\documentclass[aps,prl,reprint,groupedaddress]{revtex4-1}
\usepackage{graphicx} % Include figure files
\usepackage{dcolumn} % Align table columns on decimal point
\usepackage{bm}% bold math
\usepackage{hyperref}% add hypertext capabilities

\begin{document}
\title{Collective flows of clusters and pions in heavy-ion collisions at GeV energies }
\author{Heng-Jin Liu}
\author{Hui-Gan Cheng}
\author{Zhao-Qing Feng}
\email{Corresponding author: fengzhq@scut.edu.cn}
\affiliation{School of Physics and Optoelectronics, South China University of Technology, Guangzhou 510640, China }
\date{\today}

\begin{abstract}
Within the framework of the quantum molecular dynamics transport model, the collective flows of clusters and pions in heavy-ion collisions have been systematically investigated. The clusters are recognized by the Wigner phase-space density approach at the stage of freeze out in nuclear collisions, i.e., deuteron, triton, $^{3}$He and $\alpha$. The directed and elliptic flows of protons and deuterons in the reaction of $^{197}$Au+$^{197}$Au at incident energy 1.23\emph{A} GeV are nicely consistent with the recent HADES data. The higher order collective flows, i.e., triangular and quadrangle flows, manifest the opposite trends with the less amplitude in comparison with the rapidity distributions of directed and elliptic flows. The flow structure of $^{3}$He and $\alpha$ is very similar to the proton spectra. The influence of the pion potential on the pion production is systematically investigated and compared with the FOPI data via the transverse momentum, longitudinal rapidity and collective flows in collisions of $^{197}$Au + $^{197}$Au. It is manifested that the pion yields are slightly suppressed in the domain of mid-rapidity and high momentum. The antiflow phenomena is reduced by implementing the pion potential and more consistent with the FOPI data in collisions of $^{197}$Au+$^{197}$Au at the incident energy 1.5\emph{A} GeV.

\begin{description}
\item[PACS number(s)]
24.10.Lx, 25.70.Mn, 25.75.Ld
\end{description}
\end{abstract}

\maketitle

\section{I. Introduction}
Cluster and particle production in heavy-ion collisions at intermediate energies bring the in-medium information in dense matter, i.e., the correlation and binding of a few of nucleons, cluster potential in nuclear matter, in-medium properties of particles etc. The collective flows manifest the anisotropy emission of particles from the fireball formed in heavy-ion collisions \cite{Ol92}. The in-medium properties of hadrons and dense nuclear matter might be extracted from the analysis of collective flows, i.e., the rapidity, transverse momentum, kinetic energy spectra etc. The collective flows of nuclear fragments, free nucleons, pions, kaons and hyperons have been extensively investigated both in theories and in experiments \cite{Bu84,Gu84,Ki97,Ra99,An01,Ch01,Da02,Re07,Fe10,He13}.
Heavy-ion collisions at GeV energies provide the possibility for investigating strongly interacting matter with the high temperature and high-baryon density in laboratories, which are significant for extracting the properties of nuclear equation of state (EOS), the first-order phase transition of quark-glue matter and hadronic matter, and for searching the critical point. The collective flows manifest the decomposition of the azimuthal angle $\phi$ distribution of emitted particles from the Fourier expansion with respect to the reaction plane \cite{Vo96}. The in-plane and out-of-plane flows of particle emission are usually investigated to extract the information of early stage in heavy-ion collisions, in particular the flow magnitude, slope parameter and flow structure in phase space \cite{So96,Re97,He99,An06,Hi18,Hi20,Da22}.
In recent years, the higher order flow harmonics of protons, deuterons and tritons in collisions of
$^{197}$Au + $^{197}$Au at the invariant energy of $\sqrt{s_{NN}}$=2.4 GeV were measured by the HADES collaboration \cite{Ad20, Ad22}. Calculations with transport models gave that the triangular flow coefficient $v_{3}$ exhibits the enhanced sensitivity to the nuclear equation of state in heavy-ion collisions at several GeV energies and a nonvanishing fourth-order coefficient $v_{4}$ can be used to constrain the nuclear mean-field potential at high-baryon densities \cite{Hi18,Hi20}.

On the other hand, the collective flows of hyperons and hypernuclides are promising observables for investigating the hypernuclear formation, hyperon-nucleon interaction potential and hyperon-nucleon scattering \cite{Fe21}. The
azimuthal distribution of pions in heavy-ion collisions is influenced by the pion-nucleon potential, delta-nucleon interaction, rescattering cross sections of pions and nucleons and stiffness of symmetry energy. The deep analysis of pion flows is also helpful for shedding light on the uncertainties of constraining the high-density symmetry energy, which has important application in determining the mass-radius relation and the maximal mass of neutron stars, and is associated with the binary merging of neutron stars, frequency of gravitational wave signal, X-ray and neutrino spectra from neutron stars \cite{La23}.

In this work, the collective flows of nuclear clusters and pions in collisions of $^{197}$Au+$^{197}$Au at 1.23\emph{A} GeV and 1.5\emph{A} GeV are systematically investigated by the Lanzhou quantum molecular dynamics (LQMD) transport model. The experimental data from the HADES and FOPI collaborations are thoroughly analyzed. The article is organized as follows. In Sec. \uppercase\expandafter{\romannumeral2} we briefly introduce the LQMD model and light fragment recognition. The results are discussed in Sec. \uppercase \expandafter{\romannumeral3}. A summary and an outlook are given in Sec. \uppercase\expandafter{\romannumeral4}.

\section{II. The model description }

In the LQMD transport model, the production of resonances, hyperons and mesons is coupled in the reactions
of meson-baryon and baryon-baryon collisions, which has been used for the nuclear dynamics in heavy-ion collisions and hadron induced reactions \cite{Fe11,Fe18}. The temporal evolutions of nucleons and nucleonic resonances are described by Hamilton's equations of motion under the self-consistently generated two-body and three-body interaction potentials with the Skyrme force for the $i-$th nucleon in the system as
\begin{equation}
\dot{\textbf{r}_{i} }  =  \frac{\partial H}{\partial \textbf{p} _{i} }, \quad
\dot{\textbf{p}_{i} } =  -\frac{\partial H}{\partial \textbf{r} _{i} }.
\end{equation}
The Hamiltonian of baryons consists of the relativistic energy, Coulomb interaction, momentum dependent potential energy and local interaction as follows
\begin{equation}
H_{B}=\sum_{i}\sqrt{\textbf{p}_{i}^{2}+m_{i}^{2}}+U_{Coul}+U_{mom}+U_{loc}.
\end{equation}
Here the $\textbf{p}_{i}$ and $m_{i}$ represent the momentum and the mass of the baryons. The local interaction potential is evaluated from the energy-density functional of
\begin{equation}
U_{loc}=\int V_{loc}(\rho(\mathbf{r}))d\mathbf{r}
\end{equation}
 with
\begin{eqnarray}
V_{loc}(\rho) && = \frac{\alpha}{2}\frac{\rho^{2}}{\rho_{0}} +
\frac{\beta}{1+\gamma}\frac{\rho^{1+\gamma}}{\rho_{0}^{\gamma}} + E_{sym}^{loc}(\rho)\rho\delta^{2}
\nonumber \\
&& + \frac{g_{sur}}{2\rho_{0}}(\nabla\rho)^{2} + \frac{g_{sur}^{iso}}{2\rho_{0}}[\nabla(\rho_{n}-\rho_{p})]^{2},
\end{eqnarray}
where the $\rho_{n}$, $\rho_{p}$ and $\rho=\rho_{n}+\rho_{p}$ are the neutron, proton and total densities, respectively, and the $\delta=(\rho_{n}-\rho_{p})/(\rho_{n}+\rho_{p})$ being the isospin asymmetry of baryonic matter. The parameters $\alpha$, $\beta$, $\gamma$, $g_{sur}$, $g_{sur}^{iso}$ and $\rho_{0}$ are set to be the values of -215.7 MeV, 142.4 MeV, 1.322, 23 MeV fm$^{2}$, -2.7 MeV fm$^{2}$ and 0.16 fm$^{-3}$, respectively. The set  of the parameters gives the compression modulus of K=230 MeV for isospin symmetric nuclear matter at the saturation density ($\rho_{0}=0.16$ fm$^{-3}$). The surface coefficients $g_{sur}$ and $g_{sur}^{iso}$ are taken to be 23 MeV fm$^{2}$ and -2.7 MeV fm$^{2}$, respectively. The third term contributes the symmetry energy being of the form $E_{sym}^{loc}=\frac{1}{2}C_{sym}(\rho/\rho_{0})^{\gamma_{s}}$. The parameter $C_{sym}$ is taken as the value of 52.5 MeV. The $\gamma_{s}$ could be adjusted to get the suitable case from constraining the isospin observables, e.g., the values of 0.3, 1 and 2 being the soft, linear and hard symmetry energy and corresponding to the slope parameters of $[L(\rho _{0}) = 3\rho_{0} dE_{sym}(\rho)/d\rho|_{\rho=\rho_{0}}]$ of 42, 82, and 139 MeV, respectively. Here, the linear symmetry energy is taken into account in the calculation. Combined the kinetic energy from the isospin difference of nucleonic Fermi motion, the three kinds of symmetry energy cross at the saturation density with the value of 31.5 MeV\cite{Fe11}.

A Skyrme-type momentum-dependent potential energy is used as follows
\begin{eqnarray}
 U_{mom} && = \frac{1}{2\rho_{0}}\sum_{i,j,j \ne i}\sum_{\tau ,\tau^{\prime}} C_{\tau,\tau^{\prime}} \delta_{\tau,\tau_{i}} \delta_{\tau^{\prime}, \tau_{j}} \int\int\int d\textbf{p}d\textbf{p}^{\prime }d\textbf{r}    \nonumber\\
&& \times f_{i}(\textbf{r},\textbf{p},t) [\ln(\epsilon (\textbf{p}-\textbf{p}^{\prime})^{2}+1)^{2}] f_{j}(\textbf{r},\textbf{p}^{\prime},t).
\end{eqnarray}
Combined with Eq. (4), one can obtain the density, isospin and momentum dependent single-nucleon potential as
\begin{eqnarray}
U_{\tau}(\rho,\delta,\textbf{p}) && = \alpha\left(\frac{\rho}{\rho_{0}}\right) + \beta \left(\frac{\rho}{\rho_{0}}\right)^{\gamma} + E_{sym}^{loc}(\rho)\delta^{2}        \nonumber \\
&&  +  \frac{\partial E_{sym}^{loc}(\rho)}{\partial\rho}\rho\delta^{2} + E_{sym}^{loc}(\rho)\rho\frac{\partial\delta^{2}}{\partial\rho_{\tau}}   \nonumber \\
&&  + \frac{1}{\rho_{0}}C_{\tau,\tau} \int d\textbf{p}' f_{\tau}(\textbf{r},\textbf{p})[\ln(\epsilon(\textbf{p}-\textbf{p}')^{2}+1)]^{2}         \nonumber \\
&&  + \frac{1}{\rho_{0}}C_{\tau,\tau'} \int d\textbf{p}' f_{\tau'}(\textbf{r},\textbf{p})      \nonumber \\
&&  \times [\ln(\epsilon(\textbf{p}-\textbf{p}')^{2}+1)]^{2}.
\end{eqnarray}
Here $\tau\neq\tau'$, $\partial\delta^{2}/\partial\rho_{n}=4\delta\rho_{p}/\rho^{2}$ and $\partial\delta^{2}/\partial\rho_{p}=-4\delta\rho_{n}/\rho^{2}$. The nucleon effective (Landau) mass in nuclear matter of isospin asymmetry $\delta=(\rho_{n}-\rho_{p})/(\rho_{n}+\rho_{p})$ with $\rho_{n}$ and $\rho_{p}$ being the neutron and proton density, respectively, is calculated through the potential as $m_{\tau}^{\ast}=m_{\tau}/ \left(1+\frac{m_{\tau}}{|\textbf{p}|}|\frac{dU_{\tau}}{d\textbf{p}}|\right)$ with the free mass $m_{\tau}$ at Fermi momentum $\textbf{p}=\textbf{p}_{F}$.
Here, $C_{\tau ,\tau} = C_{mom}(1 + x), C_{\tau ,\tau^{\prime}} = C_{mom}(1-x) (\tau \ne \tau^{\prime})$ and the isospin symbols $\tau$ and $\tau^{\prime}$ represent proton or neutron, respectively. The parameters $C_{mom}$ and $\epsilon$ were determined by fitting the real part of optical potential as a function of incident energy from the proton-nucleus elastic-scattering data. In the calculation, we take the values of 1.76 MeV and 500 $c^{2}/GeV^{2}$ for $C_{mom}$ and $\epsilon$, respectively, which result in an effective mass ratio $m^{*}/m = 0.75$ in the nuclear media at saturation density for symmetric nuclear matter. The parameter $x$ as the strength of the isospin splitting with the value of -0.65 is taken in this work, which has the mass splitting of $m^{*}_{n} > m^{*}_{p}$ in the nuclear medium.

\subsection{2.1 light fragment recognition}
For the light fragment with $Z\leq$2, the Wigner phase-space density is used to evaluate the probability of fragment formation. It is assumed that the cold clusters are created at the freeze-out stage in heavy-ion collisions. The momentum distribution of a cluster with $M$ nucleons and $Z$ protons for a system with $A$ nucleons is given by
\begin{eqnarray}
\frac{dN_{M}}{d\textbf{P}} &&= G_{M}{A \choose M} {M \choose Z}\frac{1}{A^{M}}\int \prod_{i=1}^{Z}f_{p}(\textbf{r}_{i},\textbf{p}_{i}) \prod_{i=Z+1}^{M}f_{n}(\textbf{r}_{i},\textbf{p}_{i})                  \nonumber \\
&& \times \rho^{W}(\textbf{r}_{k_{1}},\textbf{p}_{k_{1}},...,\textbf{r}_{k_{M-1}},\textbf{p}_{k_{M-1}})             \nonumber \\
&& \times \delta(\textbf{P}-(\textbf{p}_{1}+...+\textbf{p}_{M}))d\textbf{r}_{1}d\textbf{p}_{1}...d\textbf{r}_{M}d\textbf{p}_{M}.
\end{eqnarray}
Here the $f_{n}$ and $f_{p}$ are the neutron and proton phase-space density, which are obtained by performing Wigner transformation based on Gaussian wave packet. The relative coordinate $\textbf{r}_{k_{1}}, ..., \textbf{r}_{k_{M-1}}$ and momentum $\textbf{p}_{k_{1}}, ..., \textbf{p}_{k_{M-1}}$ in the $M-$nucleon rest frame are used for calculating the Wigner density $\rho^{W}$ \cite{Ma97,Ch03}. The spin-isospin statistical factor $G_{M}$ is 3/8, 1/12 and 1/96 corresponding to M=2, 3 and 4, respectively. The root-mean-square radii of intending a cluster is needed for the Wigner density, i.e., 1.61 fm and 1.74 fm for triton and $^{3}$He. The method is also extended to recognize the hypernuclide production in heavy-ion collisions \cite{Fe20}. It should be noticed that the nuclear structure effect is neglected by the method and the $\alpha$ yields are strongly underestimated in comparison with the FOPI data \cite{Re10}. The clusters can be also created in nucleon-nucleon or nucleon-cluster collisions, which are determined by the correlation of nucleons and are associated with the nuclear density and nucleon momentum \cite{On19}. The cluster production at the freeze-out stage in nuclear collisions is also contributed from the de-excitation of primary fragments \cite{Fe20}, in particular in the low-energy cluster yields.
The primary fragments are constructed in phase space with a coalescence model, in which the nucleons at
freeze-out stage are considered to belong to one fragment with the relative momentum smaller than $P_{0}$ and with the relative distance smaller than $R_{0}$ (here $P_{0}$ = 200 MeV/c and $R_{0}$ = 3 fm). The secondary fragments are estimated with the help of the GEMINI++ code via the sequential binary decay by Charity \cite{Ch10}.

\subsection{2.2 Pion production in heavy-ion collisions}

At the near threshold energy, the production of pions are mainly contributed from the direct process and decay of the resonances $\Delta$(1232), $N^{*}$(1440) and $N^{*}$(1535). The relation channels are given as follows
\begin{eqnarray}
&&    N N\leftrightarrow N\Delta , \quad   NN\leftrightarrow NN^{*}, \quad  NN\leftrightarrow \Delta \Delta,   \nonumber \\
&& \Delta \leftrightarrow N\pi,  \ N^{*}\leftrightarrow N\pi,   NN\to NN\pi (s-state).
\end{eqnarray}
The cross sections of each channel to produce resonances are parameterized by fitting the experimental data calculated with the one-boson exchange model \cite{Hu94}. The energy- and momentum- dependent decay width is used in the calculation. The probabilities of specific decay channels are as follows \cite{Ts99}
\begin{eqnarray}
\Delta ^{+} \leftrightarrow \frac{1}{3} \left ( \pi ^{+} +  n \right )+  \frac{2}{3} \left ( \pi ^{0}+p\right ),    \nonumber \\
\Delta ^{0} \leftrightarrow \frac{1}{3} \left ( \pi ^{-} +  p \right )+  \frac{2}{3} \left ( \pi ^{0}+n\right ),    \nonumber \\
\Delta ^{- } \leftrightarrow 1\left(n+  \pi ^{- }\right) ,  \Delta ^{++} \leftrightarrow 1\left(p+  \pi ^{+}\right).
\end{eqnarray}
The coefficient of branching ratio is determined by the square of the Clebsch-Gordan coefficients.

The transportation of pion in nuclear medium is also given by
\begin{equation}
    H_{M} = \sum_{i=1}^{N_{M} } \left [ V_{i}^{Coul} +  \omega ( \textbf{p}_{i}, \rho _{i} )  \right ].
\end{equation}
The Coulomb interaction is given by
\begin{equation}
    V_{i}^{Coul}= \sum_{j=1}^{N_{B} } \frac{e_{i}e_{j} }{r_{ij} },
\end{equation}
where the $N_M$ and $N_B$ are the total numbers of mesons and baryons including charged resonances, respectively. It should be noted that the pion meson is taken as the point particle and the Coulomb interaction between mesons is neglected owing to the limited numbers in comparison with the baryons.

The pion energy in the nuclear medium is composed of the isoscalar and isovector contributions as
\begin{equation}
    \omega _{\pi }( \textbf{p}_{i}, \rho _{i} ) = \omega _{isoscalar} ( \textbf{p}_{i}, \rho _{i} ) +  C_{\pi }\tau _{z}\delta \left ( \rho /\rho _{0}  \right )^{\gamma _{\pi } }.
\end{equation}
The coefficient $C_{\pi}$ = $\rho_{0}\hbar^{3}/(4f^{2}_{\pi})$ = 36 MeV, and the isospin quantities are taken as $\tau _{z}$= -1, 0, and 1 for $\pi^{+}$, $\pi^{0}$, and $\pi^{-}$, respectively \cite{Fe15}. The isospin asymmetry $\delta$=($\rho_n - \rho_p$)/($\rho_n + \rho_p$ ) and the quantity $\gamma_{\pi}$ adjusts the isospin splitting of the pion optical potential. We take $\gamma_{\pi}$=2 in the model. The pion potential is estimated by the in-medium energy via the relation $V_{\pi}^{opt} (\textbf{p},\rho) = \omega_{\pi} (\textbf{p},\rho) - (m_{\pi}^{2}+\textbf{p}^{2})^{1/2} $. The pion-nucleus scattering and charge exchange reactions were nicely explained with the potential to some extent \cite{Fr07,Fe16}.

For the evaluation of the isoscalar part, we chose $\Delta$-hole model \cite{Br75, Fr81} which is given by
\begin{eqnarray}
    \omega _{isoscalar} \left ( p_{i} ,\rho _{i}  \right ) && =  S_{\pi }  \left ( p_{i} ,\rho _{i} \right )\omega _{\pi - like} \left (p_{i} ,\rho _{i}  \right )    \nonumber \\
   && + S_{\Delta }  \left ( p_{i} ,\rho _{i} \right )\omega _{\Delta  - like} \left (p_{i} ,\rho _{i}  \right ).
\end{eqnarray}
The probability of the pion component satisfies the relation by
\begin{equation}
    S_{\pi }  \left ( p_{i} ,\rho _{i} \right ) + S_{\Delta }  \left ( p_{i} ,\rho _{i} \right ) = 1.
\end{equation}
The value of the probability is determined from the pion self-energy as \cite{Xi93}
\begin{equation}
    S\left ( p_{i} ,\rho _{i} \right ) = \frac{1}{1-\partial\Pi \left ( \omega  \right )/\partial \omega ^{2}   },
\end{equation}
where the pion self-energy is given by
\begin{equation}
    \Pi =p_{i}^{2}\frac{\chi }{1-{g}' \chi } ,
\end{equation}
 the Migdal parameter ${g}'\sim$ 0.6 and
 \begin{equation}
     \chi =- \frac{8}{9} \left ( \frac{f_{\Delta } }{m_{\pi } }  \right )^{2}  \frac{\omega _{\Delta }\rho \hbar ^{3}  }{\omega _{\Delta }^{2}-\omega ^{2}  }exp\left ( -2p_{i}^{2} /b^{2}  \right ).
 \end{equation}
 $\omega _{\Delta }$ = $(m_{\Delta}^{2} + p_{i}^{2})^{1/2} - m_{N}$, $m_{\pi}$, $m_{N}$, and $m_{\Delta}$ are the pion, nucleon, and delta masses, respectively. The $\pi N \Delta$ coupling constant $f_{\Delta} \sim$ 2 and the cutoff factor b $\sim$ 7$m_{\pi}$. Two eigenvalues of $\omega_{\pi-like}$ and  $\omega_{\Delta-like}$ are obtained from the pion dispersion relation as
 \begin{equation}
     \omega ^{2}   = p_{i}^{2} + m_{\pi }^{2} + \Pi \left ( \omega  \right ).
 \end{equation}

The $\Delta$-nucleon interaction is estimated via the nucleon optical potential by
 \begin{eqnarray}
  && U_{\Delta ^{- } }  =U_{n}, \quad   U_{\Delta ^{++ } }  =U_{p},  \quad
  U_{\Delta ^{+ }} = \frac{1}{3} U_{n} + \frac{2}{3}U_{p},      \nonumber \\
  &&  U_{\Delta ^{0 } }  =\frac{1}{3} U_{p}+  \frac{2}{3}U_{n},
\end{eqnarray}
 where the $U_{n}$ and $U_{p}$ are the single-particle potentials for neutron and proton in Eq. (6), respectively. The N$^{\ast}$-nucleon potential is taken as the same with the nucleon-nucleon potential.

%%%%%%%%%%%%%%%%%%%%%%%%%%%%%%%%%%%% figure 1 %%%%%%%%%%%%%%%%%%%%%%
\begin{figure*}
 \includegraphics[height=6.5cm,width=8cm]{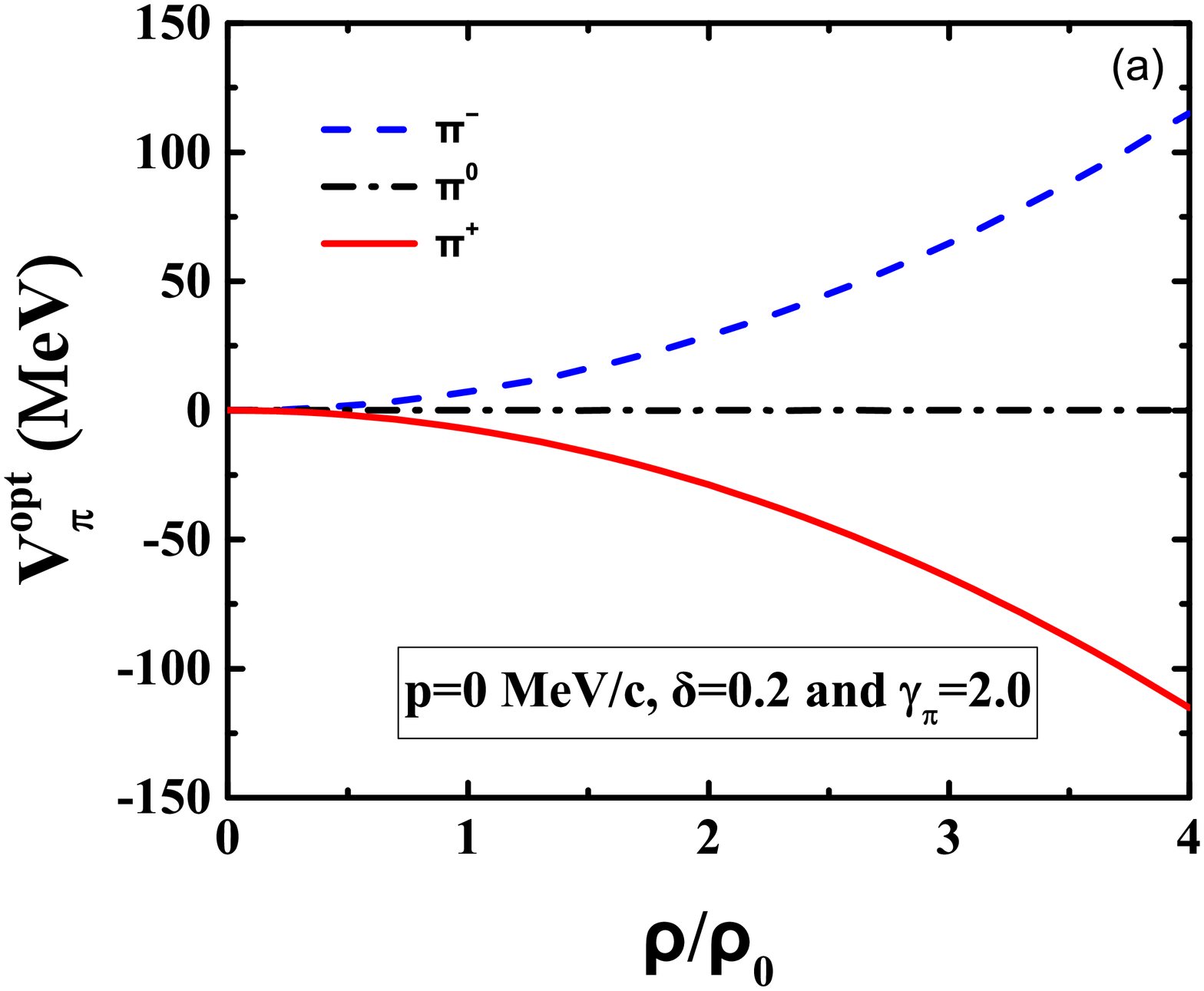}    \includegraphics[height=6.5cm,width=8cm]{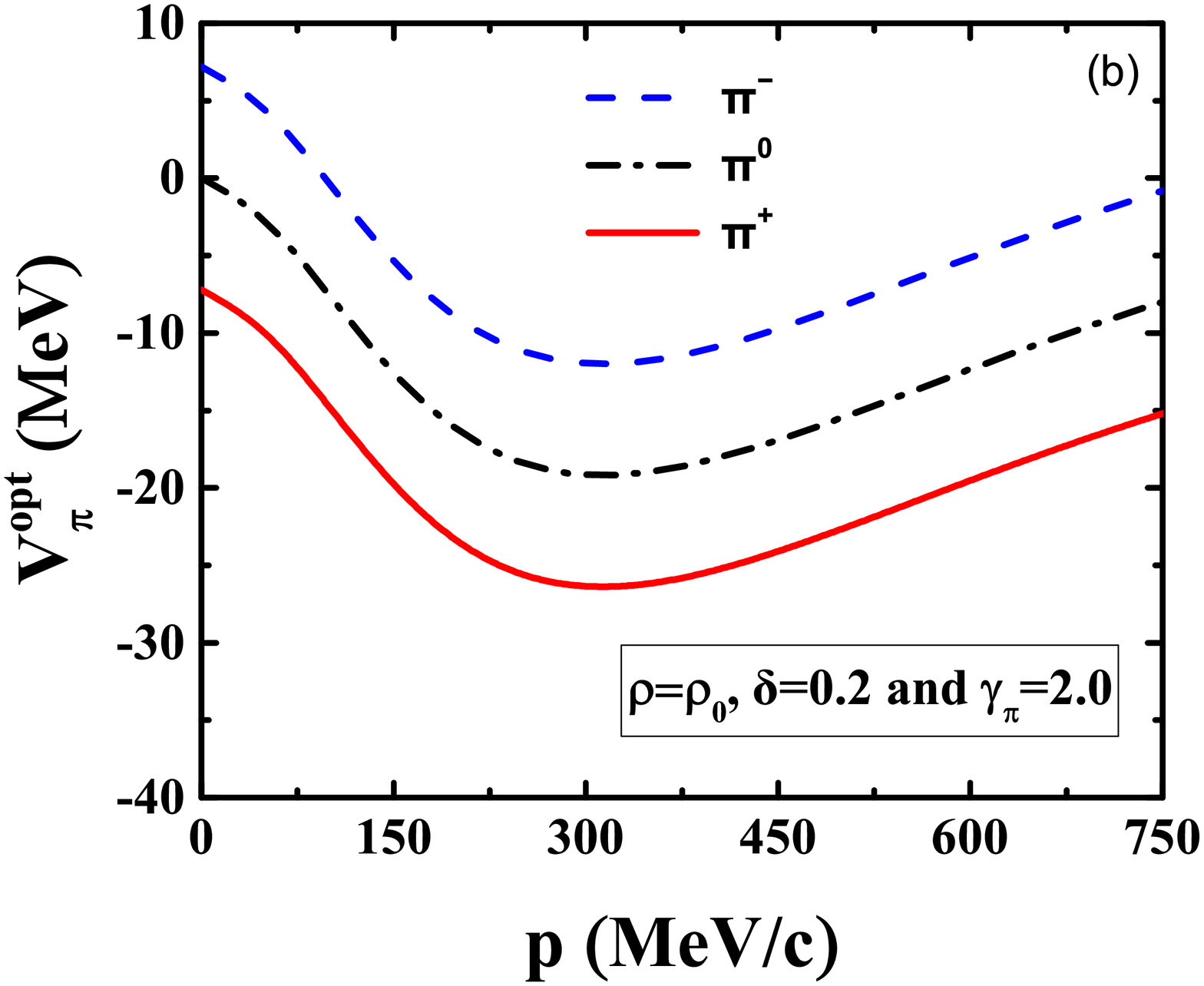}
 \caption{ (a) Density and (b) momentum dependence of the optical potential of pion in the neutron-rich nuclear matter with the isospin asymmetry $\delta=$0.2. }
\end{figure*}
%%%%%%%%%%%%%%%%%%%%%%%%%%%%%%%%%%%%%%%%%%%%%%%%%%%%%%%%%%%%%%%

 The energy balance in the decay of resonances and reabsorption of pion in nuclear medium is satisfied by the relation
 \begin{eqnarray}
 \sqrt{m_{R}^{2}+\textbf{p}_{R}^{2}} + U_{R}(\rho,\delta,\textbf{p}) && = \sqrt{m_{N}^{2} + \left (\textbf{p}_{R} - \textbf{p}_{\pi } \right )^{2} } + U_{N}(\rho,\delta,\textbf{p})          \nonumber \\
&&  + \omega _{\pi} \left (\textbf{p}_{\pi },\rho  \right)+V_{\pi N}^{Coul}.
 \end{eqnarray}
The $\textbf{p}_{R}$ and $\textbf{p}_{\pi}$ are the momenta of resonance and pion, respectively. The $U_{R}$ and $U_{N}$ are the singe-particle potentials for resonance and nucleon. The last term $V_{\pi N}^{Coul}$ has the contribution only for the charged pair channels of $\triangle^{0}\leftrightarrow  \pi^{-}+p$ and $\triangle^{++}\leftrightarrow  \pi^{+}+p$. Recently, the influence of the pion potential on pion dynamics in heavy-ion collisions has been extensively investigated with different transport models \cite{Ho14,Fe15,Gu15,So15,Co16,Zh17,Fe17}.
Shown in Fig. 1 is the pion optical potential as functions of the pion momentum and baryon density at the saturation density with the isospin asymmetry $\delta=(\rho_{n}-\rho_{p})/(\rho_{n}+\rho_{p})$. It is obvious that isospin splitting of the pion potential appears and the effect is pronounced in the domain of high baryon density, which impacts the charged-pion ratios in heavy-ion collisions. The minimum position with the momentum dependence is close to the resonance energy (p=290 MeV/c). The difference of charged pion potentials is similar to the contribution of the s-wave potential by fitting the chiral perturbation theory calculations with the positive value of $V_{\pi^{-}}^{opt} - V_{\pi^{+}}^{opt} $ \cite{Zh17}.

\subsection{2.3 Collective flows}
It is well known that the EOS and the nuclear dynamics in heavy-ion collisions have been widely studied through the analysis of collective flows, such as nucleon, light particle, and meson flows \cite{Ki97, Ra99, Fe10, An01}. The azimuthal distributions of particles produced in the intermediate energy heavy-ion collisions is conveniently parameterised by a Fourier decomposition with the coefficients $v_{n}(p_{t}, y)$ as
\begin{eqnarray}
\frac{dN}{N_{0}d\phi} \left (y,p_{t} \right) &&= 1+2v_{1}(y,p_{t}) \cos(\phi) + 2v_{2}(y,p_{t}) \cos (2\phi)    \nonumber \\
&& + 2v_{3}(y,p_{t}) \cos(3\phi) + 2v_{4}  (y,p_{t})\cos(4\phi),
\end{eqnarray}
in which $p_t=\sqrt{p_x^2+p_y^2}$ and $y=\frac{1}{2}\ln{\frac{E+p_z}{E-p_z}}$ are the transverse momentum and the longitudinal rapidity with the total energy $E$, respectively. The directed flow $v_{1}=\left \langle p_{x}/p_{t} \right \rangle =\left \langle cos(\phi) \right \rangle$ and the elliptic flow $v_{2}=\left \langle(p^{2}_{x}-p^{2}_{y})/p^{2}_{t}\right \rangle =\left \langle cos(2\phi)\right \rangle$ manifest the competition between the in-plane $(v_{2}>0)$ and out-of-plane $(v_{2}<0)$ particle emissions. The triangular flow $v_3=\left\langle\left(p_x^3-3p_xp_y^2\right)/p_t^3\right\rangle=\left\langle \cos(3\phi) \right\rangle$ and the quadrangular flow $v_4=\left\langle\left(p_x^4+p_y^4-6p_x^2p_y^2\right)/p_t^4\right\rangle=\left\langle \cos (4\phi) \right\rangle$ manifest the anisotropic distributions on the plane perpendicular to the beam direction. The directed flow in the reaction plane is influenced by the pressure gradient of nuclear matter formed heavy-ion collisions. The difference between in-plane and out-of-plane emission of particles is embodies via the elliptic flow distribution. The flow rapidity and transverse momentum distributions of particles are related to the collision centrality of reaction system, mean-field potential, scattering and reabsorption process of particle in nuclear medium, symmetry energy etc.

\section{III. Results and discussion }

 %%%%%%%%%%%%%%%%%%%%%%%%%%%%%%%%%%%% figure 2 %%%%%%%%%%%%%%%%%%%%%%
\begin{figure*}
\includegraphics[height=12cm,width=15cm]{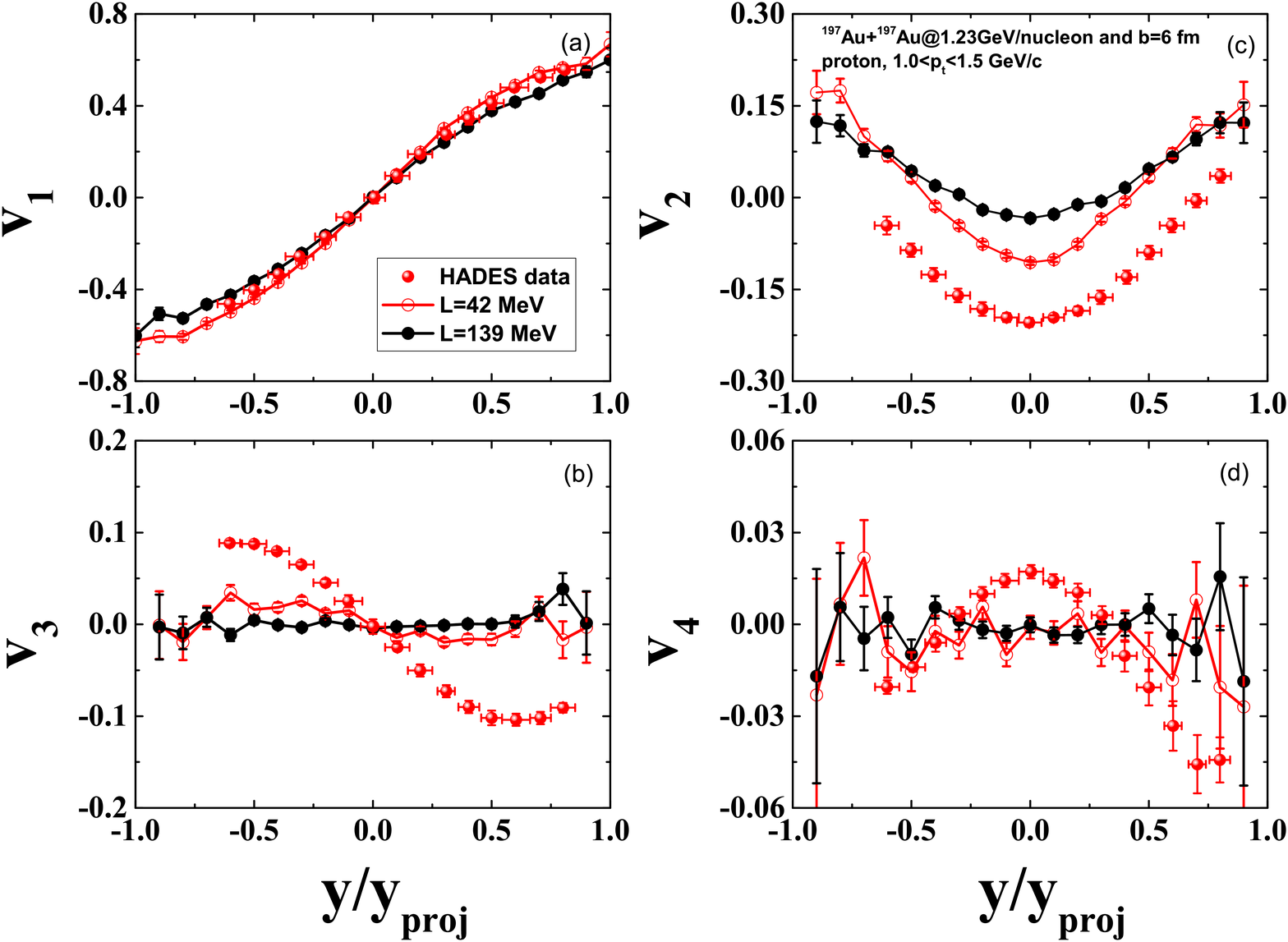}
\caption{ The rapidity distribution of the collective flows of protons in the reaction of $^{197}$Au+$^{197}$Au at incident energy 1.23\emph{A} GeV and with the collision parameter of b=6 fm. The results are averaged over the transverse momentum in the domain $1.0< p_{t}< 1.5$ GeV/c. The experimental data are taken from the HADES collaboration \cite{Ad20}. }
\label{fig.2}
\end{figure*}
%%%%%%%%%%%%%%%%%%%%%%%%%%%%%%%%%%%%%%%%%%%%%%%%%%%%%%%%%%%%%%%

 %%%%%%%%%%%%%%%%%%%%%%%%%%%%%%%%%%%% figure 3 %%%%%%%%%%%%%%%%%%%%%%
\begin{figure*}
    \includegraphics[height=12cm,width=15cm]{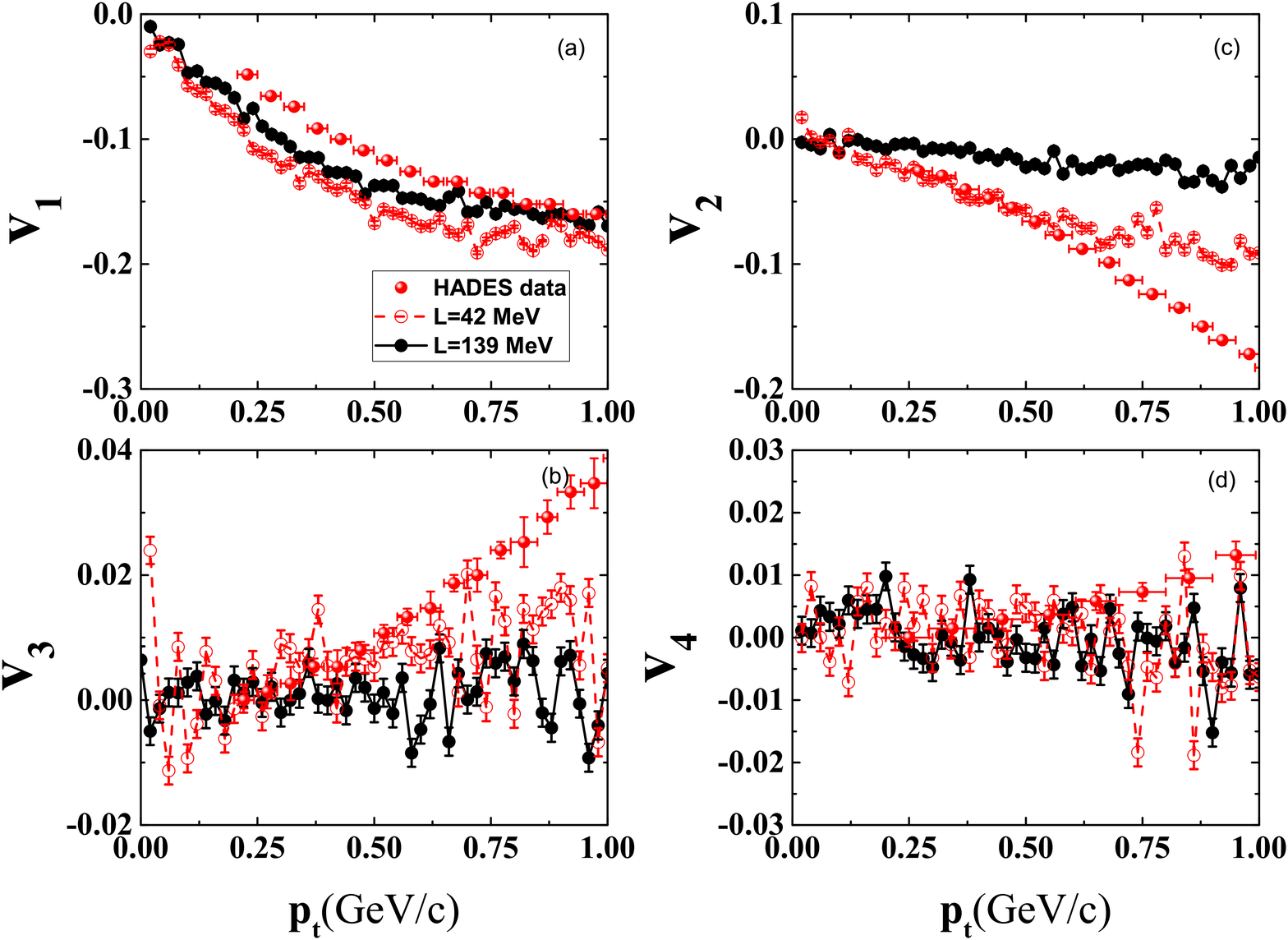}
    \caption{ The same as in Fig. 2, but for the momentum spectra in the rapidity region of $-0.25< y_{cm}< -0.15$ with the soft (L=42 MeV) and hard (L=139 MeV) symmetry energies, respectively. }
\label{fig.3}
\end{figure*}
%%%%%%%%%%%%%%%%%%%%%%%%%%%%%%%%%%%%%%%%%%%%%%%%%%%%%%%%%%%%%%%

The cluster production in heavy-ion collisions is associated with the nucleon (cluster) mean-field potential, Pauli principle, nucleon-nucleon (cluster) collisions etc, in which the structure effects influence the cluster configuration and bound system in nuclear medium. The formation of clusters is contributed from the correlation of nucleons in phase-space and also from the fragmentation process in nuclear collisions. The statistical decay of primary fragments in the spinodal reactions dominates the low-energy cluster formation (below 10 MeV/nucleon). The medium and high-energy cluster production is contributed from the coalescence approach at the final stage and partly attributed from the correlation of nucleons during the collisions \cite{Ch21}. In the past several decades, the nuclear fragmentation reactions have been extensively investigated both in experiments and in theories, in particular on the issues of spinodal multifragmentation, liquid-gas phase transition, properties of highly excited nuclei, symmetry energy at subsaturation densities etc \cite{Ch04,Co93,Po95,Wu98,Ma99}. The nuclear clusters with the charge number Z$\leq$2 manifest the baryonic matter properties in high-energy heavy-ion collisions and the bound state in the hadronization of quark-gluon plasma (QGP), might be the probes of the first order phase transition. On the other hand, the cluster spectra in phase space may shed light on the nuclear equation of state at high-baryon density. The collective flows manifest the phase space distribution of particles, in which the in-medium effect and EOS influence the flow structure. The collective flows have been measured by the HADES collaboration \cite{Ad20}. Recently, the collective flows of protons and deuterons for the HADES data were investigated by the isospin-dependent quantum molecular dynamics (IQMD) model \cite{Fa23}. We calculated the collective flows of proton and light nuclei produced in the $^{197}$Au + $^{197}$Au collisions at the incident energy of 1.23\emph{A} GeV and with the semicentral collisions with the impact parameter of b=6 fm. The collective flows of protons is given in Fig. 2 and compared with the HADES data with different symmetry energy. It is obvious that the directed flows are nicely consistent with the experimental data. The 'U' shape of elliptic flow $v_{2}$ is basically reproduced, which is caused from the out-of-plane emission of protons bounced off the reaction plane. The triangular flow $v_{3}$ exhibits an opposite trend with the structure of $v_{1}$. The amplitude of quadrangular flow $v_{4}$ is very small and almost isotropic distribution. Overall, the soft symmetry energy with the slope parameter of L=42 MeV is close to the experimental data of $v_{1}$ and $v_{2}$ spectra. The flow coefficient becomes more and more weak with increasing the flow order. The transverse momentum dependence of collective flows of protons with different symmetry energy is shown in Fig. 3. The difference between the calculations and HADES data is pronounced at the high transverse momenta, which is caused partially by neglecting the cluster construction of nucleon and cluster correlation in the preequilibrium process. The transverse momentum spectra of protons are also influenced by the Pauli principle in the evolution and nucleon-nucleon collisions. The protons are mainly recognized by the coalescence approach at the freeze-out stage (100 fm/c). The discrepancies between the HADES data and calculations, in particular the high-order flows $v_{3}$ and $v_{4}$, can not be improved by adjusting the coalescence parameters $P_{0}$ = 200 MeV/c and $R_{0}$ = 3 fm. The problems might be improved by including the correlation of multinucleon collisions in phase space and by implementing the Mott effects of clusters in nuclear medium. The works are in progress.

 %%%%%%%%%%%%%%%%%%%%%%%%%%%%%%%%%%%% figure 4 %%%%%%%%%%%%%%%%%%%%%%
\begin{figure*}
    \includegraphics[height=12cm,width=15cm]{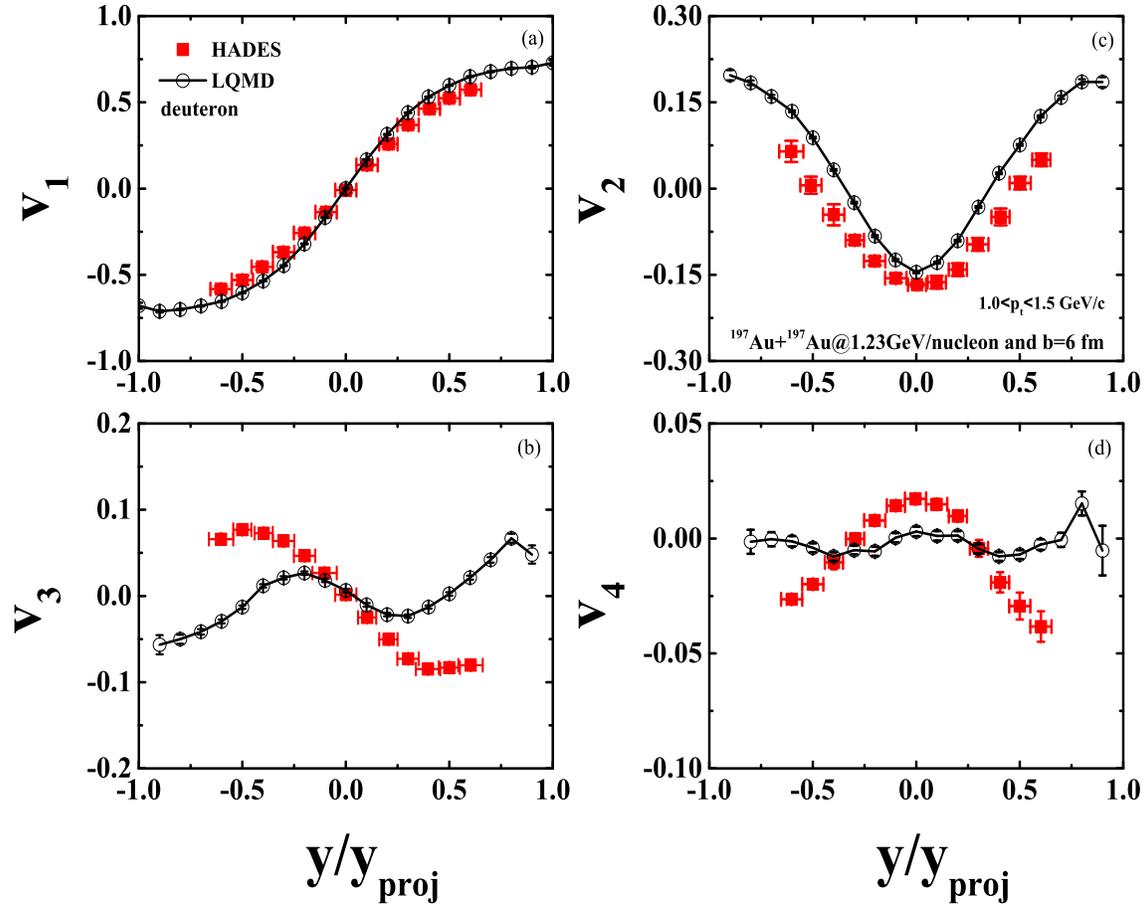}
    \caption{ Collective flows of deuterons within the transverse momentum range of $1.0< p_{t}< 1.5$ GeV/c in collisions of $^{197}$Au+$^{197}$Au at 1.23\emph{A} GeV. The experimental data are taken from the HADES collaboration \cite{Ad20}. }
\label{fig.4}
\end{figure*}
%%%%%%%%%%%%%%%%%%%%%%%%%%%%%%%%%%%%%%%%%%%%%%%%%%%%%%%%%%%%%%%

 %%%%%%%%%%%%%%%%%%%%%%%%%%%%%%%%%%%% figure 5 %%%%%%%%%%%%%%%%%%%%%%
\begin{figure*}
    \includegraphics[height=8cm,width=15cm]{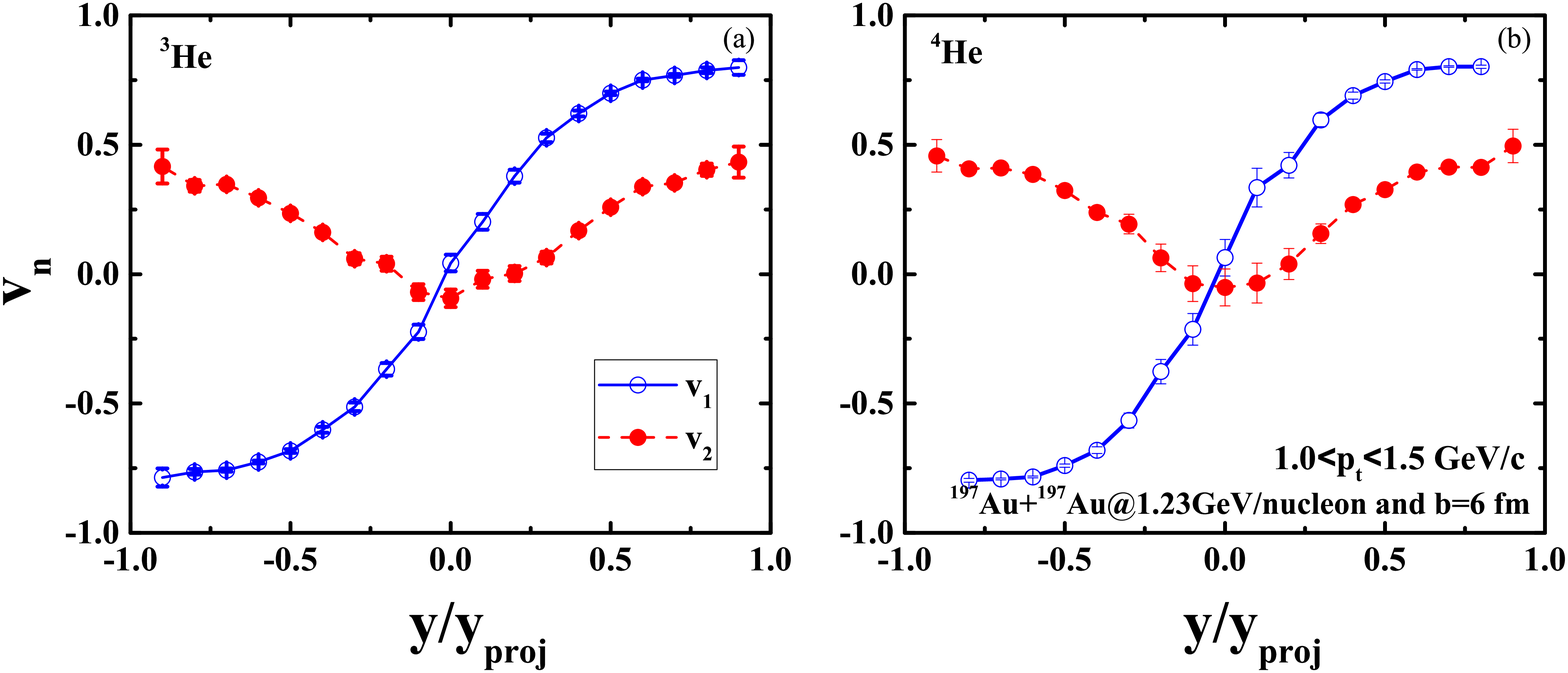}
    \caption{ Collective flows of $^{3}$He and $^{4}$He within the transverse momentum range of $1.0< p_{t}< 1.5$ GeV/c. }
\label{fig.5}
\end{figure*}
%%%%%%%%%%%%%%%%%%%%%%%%%%%%%%%%%%%%%%%%%%%%%%%%%%%%%%%%%%%%%%%

The recognition of cluster in heavy-ion collisions needs to include the bound of few-nucleon system and the Pauli principle, i.e., the construction of deuteron, triton, $^{3}$He and $^{4}$He, in which the binding energy varies with the baryon density and dissociation of cluster into nucleons in nuclear medium takes place in the evolution of reaction system. The cluster distribution in phase-space is also a sensitive probe for extracting the nuclear matter properties. Shown in Fig. 4 is the collective flows of deuterons within the transverse momentum range of $1.0< p_{t}< 1.5$ GeV/c in collisions of $^{197}$Au+$^{197}$Au at 1.23\emph{A} GeV and compared with the recent HADES data \cite{Ad20}. Similar to the proton flows, the directed and elliptic flows are nicely reproduced with the Wigner phase-space density method, namely the 'S' and 'U' shape structure. The in-plane emission of deuterons becomes obvious in the projectile-like and target-like region and the squeeze-out phenomena is pronounced in the mid-rapidity domain. The triangular flow manifests an opposite trend with the directed flow. The amplitude of quadrangular flow is very small and difficulty to reproduce the HADES data. The first and second order harmonics of anisotropic expansion are shown in Fig. 5 for the $^{3}$He and $^{4}$He production. A bit of difference with proton and deuteron emission is that the in-plane formation dominates the $^{3}$He and $^{4}$He production, which is caused that the part of cluster yields are recognized from the participating and spectator nucleons. More binding system of $^{4}$He (28.3 MeV) is available for the cluster formation, e.g., the larger yields of $^{4}$He in comparison with $^{3}$He and triton in Fermi-energy heavy-ion collisions. However, the nuclear effect becomes more and more weak with increasing the beam energy in competition with the direct multi-fragmentation, in which the cluster yields rapidly decrease with the mass number. Sophisticated investigation on the cluster formation (deuteron, triton, $^{3}$He and $^{4}$He) in nuclear collisions is in progress.

\begin{figure*}
\includegraphics[height=8cm,width=8cm]{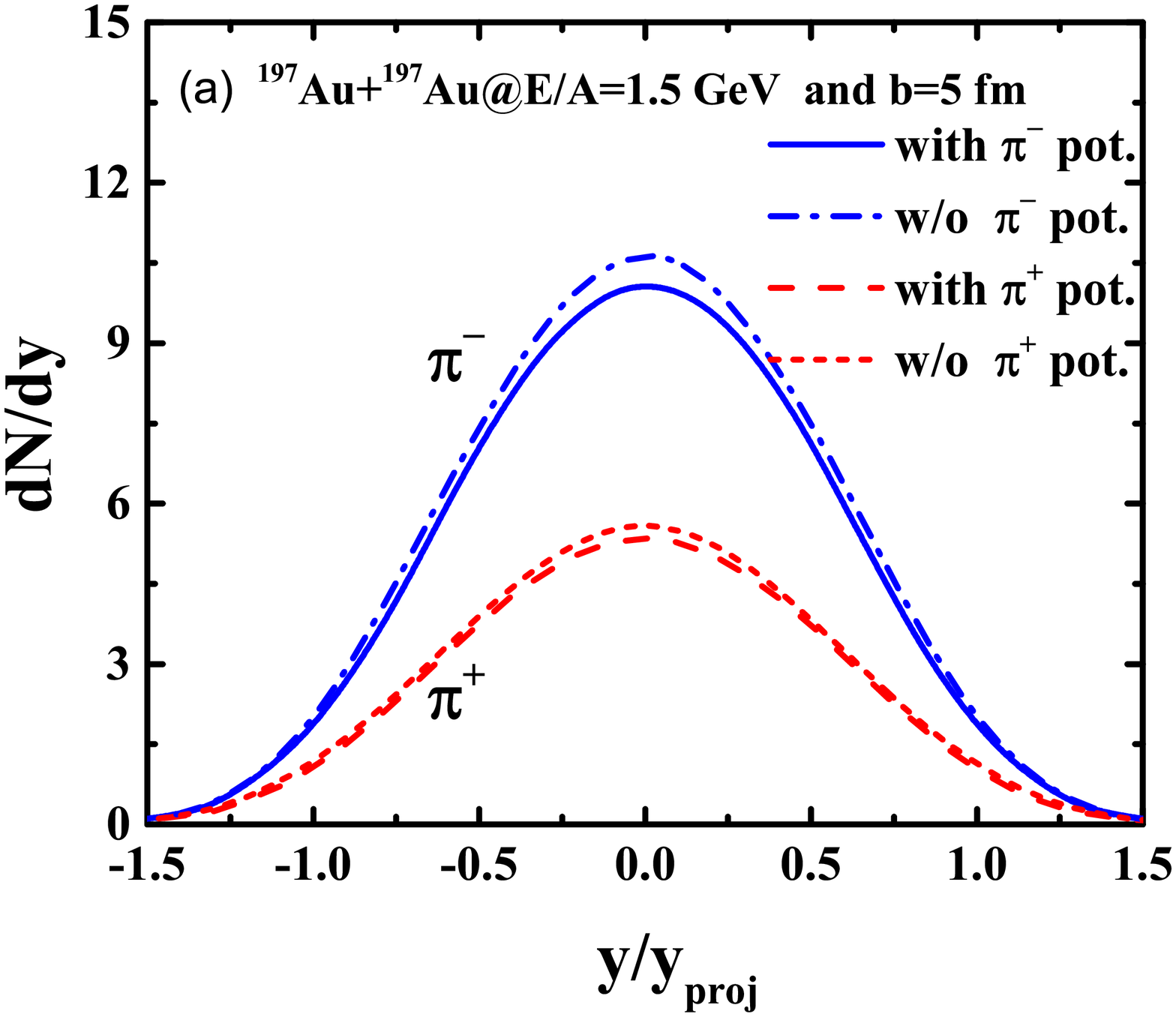}  \includegraphics[height=8cm,width=8cm]{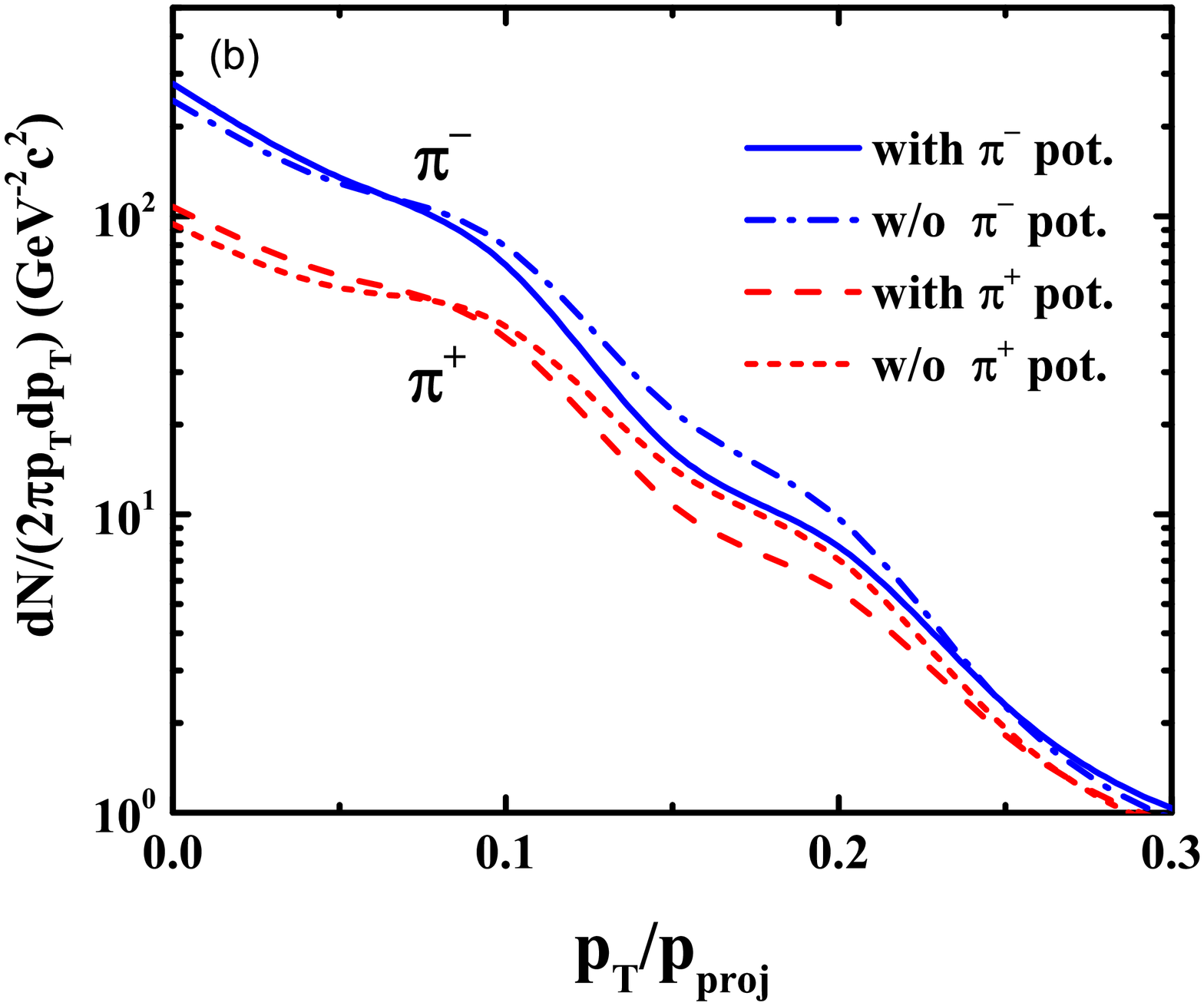}
    \caption{ (a) The rapidity and (b) transverse momentum distributions of $\pi^{-}$ and $\pi^{+}$ in $^{197}$Au + $^{197}$Au at incident energy of 1.5\emph{A} GeV. }
\label{fig.6}
\end{figure*}

Pions in heavy-ion collisions are very important for constraining the stiffness of symmetry energy. Ever since, the pion production near the threshold energy has attracted more and more attention. Before using the pion observable to extract the high-density behavior of symmetry energy, it is necessary to explore the dynamics of pions produced in heavy-ion collisions, such as rapidity distribution of yields, transverse momentum spectra, collective flows etc. At the beam energies near the threshold value (280 MeV), pions are mainly produced via the decay of the resonance $\Delta$(1232). When the pions are emitted from the reaction zone, they undergo the multiple processes $\Delta \to \pi(N) \to \Delta $. The yields are robust to be measured in experiments for constraining the high-density symmetry energy by the $\pi^{-}/\pi^{+}$ ratio. The dynamics of the pion emission calculated by transport models is helpful for the understanding on the experimental observables. We calculated the rapidity and transverse momentum distributions of pions produced in $^{197}$Au + $^{197}$Au at the incident energy of 1.5\emph{A} GeV  and at the impact parameter of b=5 fm as shown in Fig. \ref{fig.6}. The solid curves denote the results with the pion potential, and the dashed curves represent the ones without the pion potential. It is concluded that the pions are mainly produced in the mid-rapidity region, and the influence of $\pi$-N potential on charged pions is obvious. With the attractive potential, the pions are more favorable to be captured by the surrounding nucleons. The inclusion of the optical potential enhances the production of low-energy pions, but the production of high-energy pions are suppressed due to the reabsorption process. The effects of pion potential on the yields of pions gradually vanish with the increase of pion momentum. Considering the combined effects of Coulomb interaction and the optical potential characterized by its momentum dependence , it is easy to understand the transverse momentum distribution of pions. The difference of $\pi^{+}$ and $\pi^{-}$ is due to the combined effects by the optical potential and Coulomb interaction in nuclear medium.

 %%%%%%%%%%%%%%%%%%%%%%%%%%%%%%%%%%%% figure 6 %%%%%%%%%%%%%%%%%%%%%%
%%%%%%%%%%%%%%%%%%%%%%%%%%%%%%%%%%%%%%%%%%%%%%%%%%%%%%%%%%%%%%%%%%%%%

\begin{figure*}
\includegraphics[height=8cm,width=15cm]{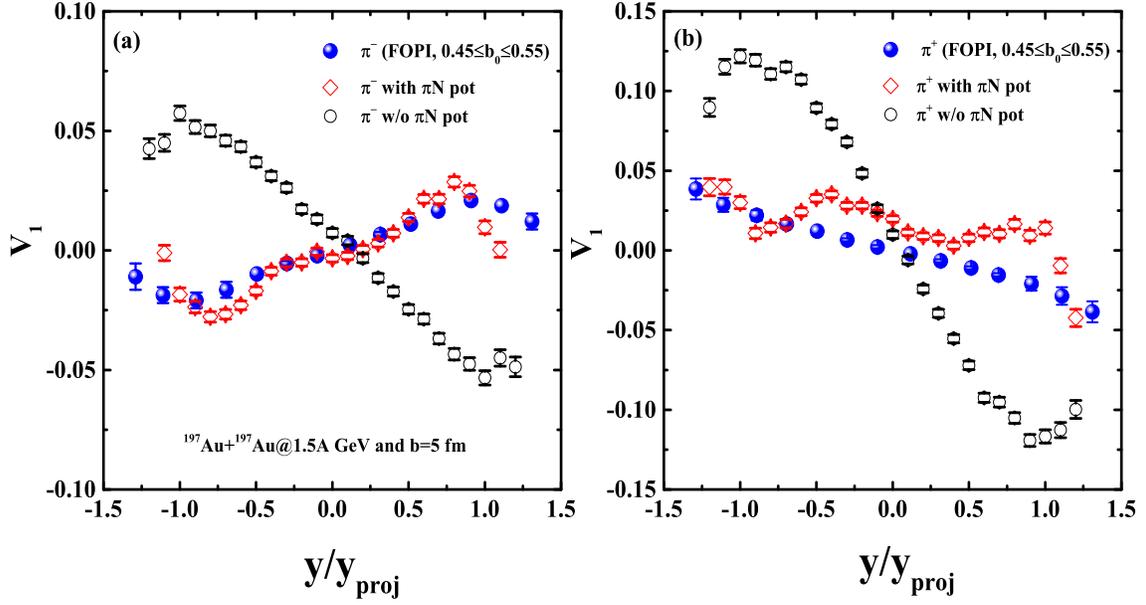}
\caption{ The directed flows of $\pi^{-}$ (a) and $\pi^{+}$ (b) in $^{197}$Au+$^{197}$Au collisions at 1.5\emph{A} GeV. The experimental data are taken from FOPI Collaboration \cite{Re07}. }
\label{fig.7}
\end{figure*}

\begin{figure*}
\includegraphics[height=8cm,width=15cm]{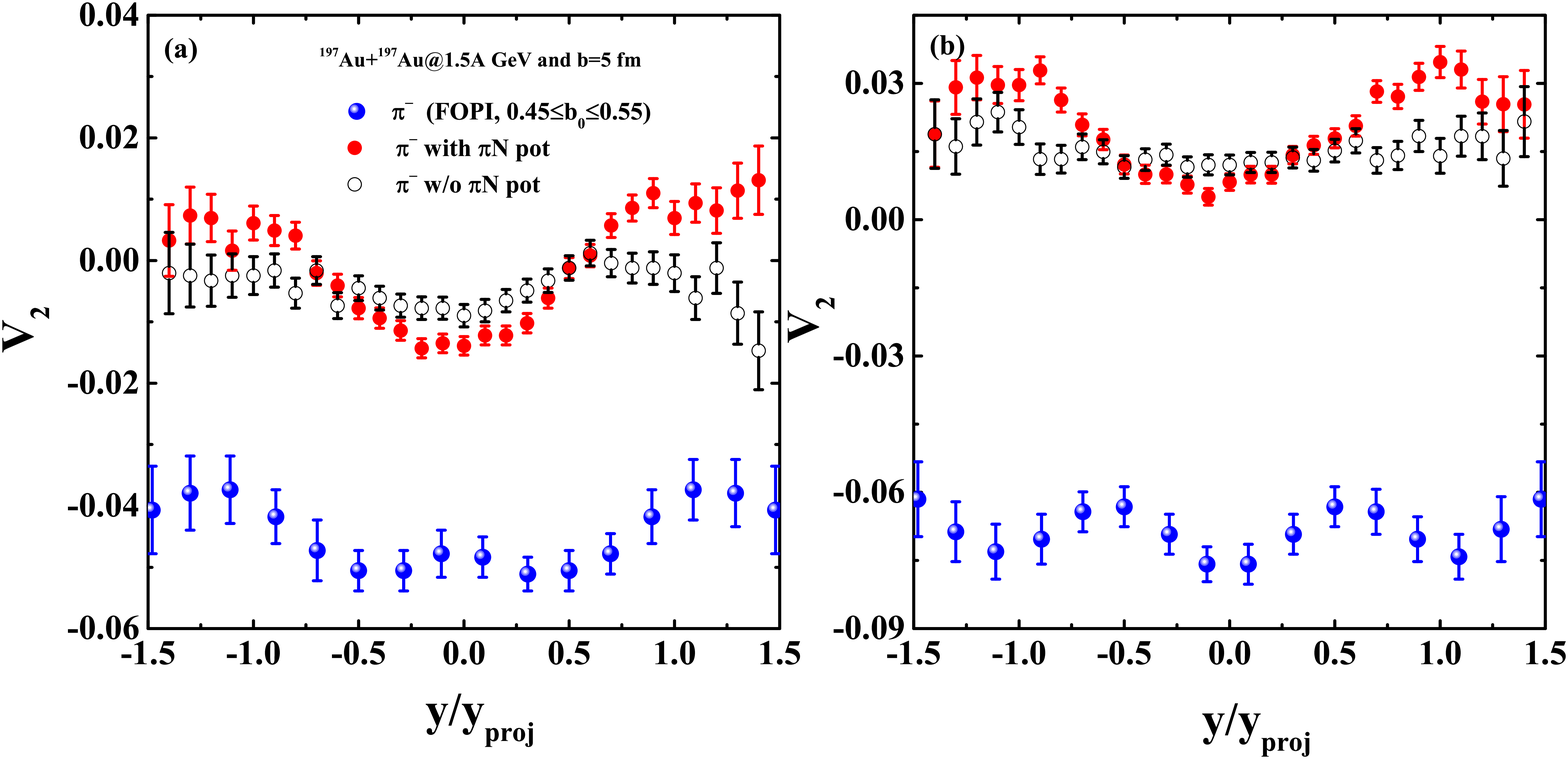}
\caption{ The same as in Fig. \ref{fig.7}, but for the elliptic flows.  }
\label{fig.8}
\end{figure*}

\begin{figure*}
\includegraphics[height=12cm,width=15cm]{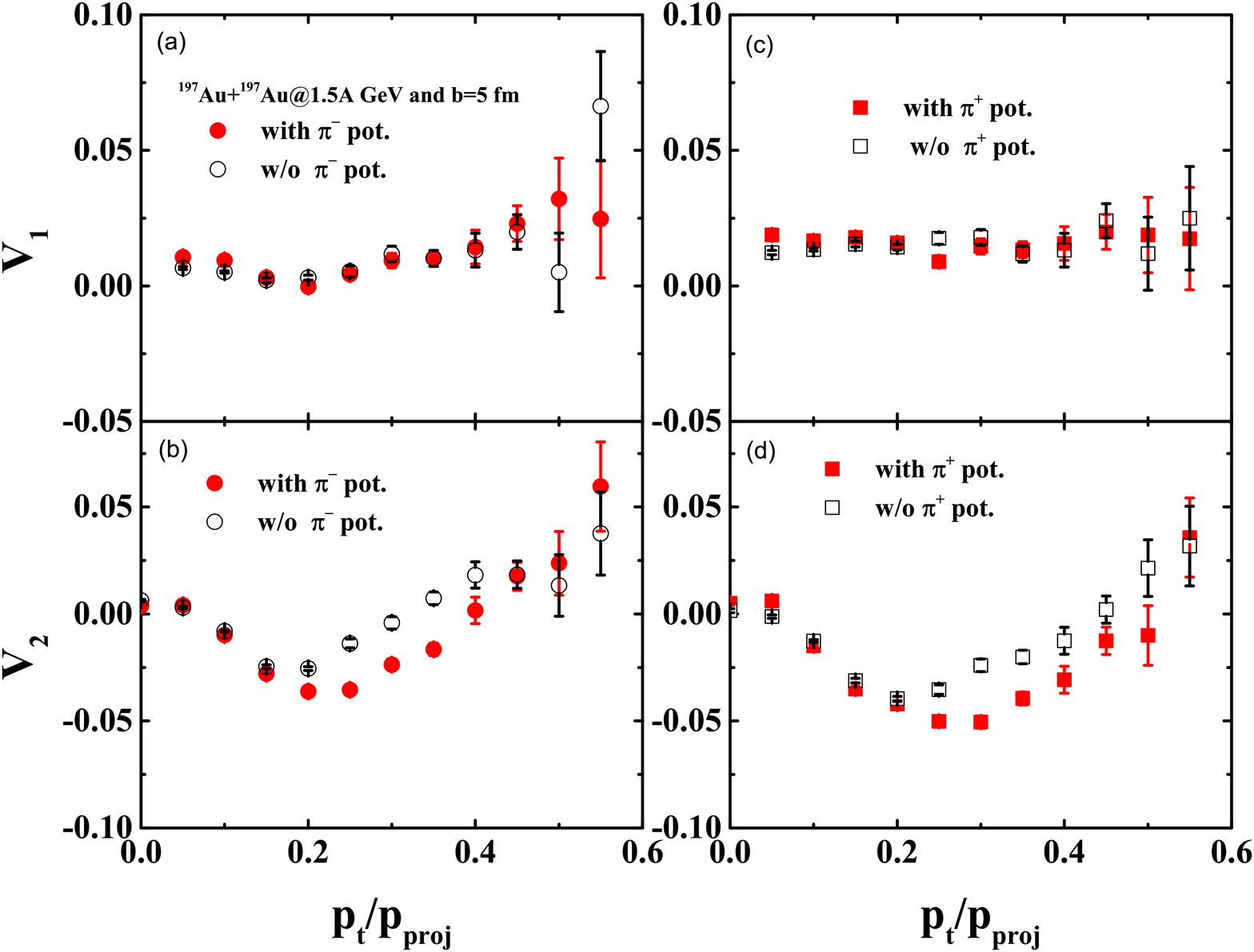}
\caption{ Comparison of the transverse momentum distributions of directed and elliptic flows with and without the pion-nucleus potential, respectively. }
\label{fig.9}
\end{figure*}

The pion mesons are usually rescattered and reabsorbed by the surrounding nucleons, in which the primordial pions are created at the compression stage in nuclear collisions. This gives rise to an apparent pion flow and offers a powerful tool for exploring the pion dynamics in nuclear medium and equation of the state of nuclear matter \cite{Ve82,Go89,Li91,Jh93}. The collective flows of pions were firstly measured by the Bevalac streamer chamber group in 800 MeV/nucleon Ne induced reactions \cite{Go89}. At the energies near the threshold value, the pions are mainly from the decay of $\Delta$(1232). So the $\Delta$-nucleon interaction potential and scattering are also important for the pion production. The emission of $\Delta$ would tend to create the similar directed flow of protons. But the rescattering and reabsorption processes of pions with the spectator nucleons lead to the appearance of antiflows of pions in comparison with the proton flows \cite{Ba95}. The collective flows manifest the phase space distribution of particles, in which the in-medium effect and EOS influences the flow structure. The pion flows have been measured and extensive investigated by the the FOPI collaboration \cite{Re07}. We calculated the directed flows of charged pions produced in the $^{197}$Au + $^{197}$Au collisions at the incident energy of 1.5\emph{A} GeV and with the semicentral collisions (b=5 fm) as shown in Fig. \ref{fig.7}. The reduced impact parameter is obtained by $b_{0}=b/1.15(A_{p}^{1/3}+A_{t}^{1/3})$ with $A_{p}$ and $A_{t}$ being the mass numbers of projectile and target nuclei, respectively. It is obvious that, when there is no pion potential, the direct flows of the charged pions both show countercurrents, as described in the preceding paragraph, but the result is close to experimental data when an attractive potential is included, as compared to the case without. The well-known S shape is apparent in the distributions of the directed flow of $\pi^{-}$. Also, the directed flow of $\pi^{+}$ shows the famous shadowing effect due to the existence of $\pi$-N potential. The difference between $\pi^{-}$ and $\pi^{+}$ is caused by the Coulomb interaction and the isospin effect. The elliptic flow in $^{197}$Au + $^{197}$Au is also investigated in the same way as in directed flow, as shown in Fig. \ref{fig.8}. Although there is a difference between our results and the data, they are the same in trend when the pion potential is included. Like in Ref. \cite{Fe10}, the flow spectra can be well reproduced in near-central collisions. However, the calculations overpredict the experimental data in peripheral collisions owing to the difference in the choices of observables for impact parameter. But it still shows that pions are out-of-plane emission. Shown in Fig. \ref{fig.9} is the transverse momentum distributions of collective flows. It is obvious that the pion potential influences the high-momentum elliptic flows both $\pi^{-}$ and $\pi^{+}$. The difference of directed flows is very small. The flow structure is similar to the FOPI data \cite{Re07}.

 %%%%%%%%%%%%%%%%%%%%%%%%%%%%%%%%%%%% figure 7 %%%%%%%%%%%%%%%%%%%%%%
%%%%%%%%%%%%%%%%%%%%%%%%%%%%%%%%%%%%%%%%%%%%%%%%%%%%%%%%%%%%%%%

 %%%%%%%%%%%%%%%%%%%%%%%%%%%%%%%%%%%% figure 8 %%%%%%%%%%%%%%%%%%%%%%
%%%%%%%%%%%%%%%%%%%%%%%%%%%%%%%%%%%%%%%%%%%%%%%%%%%%%%%%%%%%%%%

 %%%%%%%%%%%%%%%%%%%%%%%%%%%%%%%%%%%% figure 9 %%%%%%%%%%%%%%%%%%%%%%
%%%%%%%%%%%%%%%%%%%%%%%%%%%%%%%%%%%%%%%%%%%%%%%%%%%%%%%%%%%%%%%%%%%%%

\section{ Conclusion}
In summary, the collective flows of clusters and pions produced in collisions of $^{197}$Au+$^{197}$Au are thoroughly investigated within the LQMD transport model. The experimental data from the HADES and FOPI collaboration are systematically analyzed. The directed and elliptic flows of protons and deuterons are nicely reproduced and the out-of-plane emission dominates the cluster production in the rapidity domain of -0.5$<y/y_{proj}<$0.5. The triangular flow manifests the opposite structure in comparison with the directed flow but the less amplitude than the HADES data. The positive values of transverse momentum spectra of triangular and quadrangular flows are strongly underestimated. The in-plane emission of $^{3}$He and $^{4}$He dominates the cluster production. The attractive pion potential leads to the reduction of pion production in the mid-rapidity region. The directed flows of pions in the reaction of $^{197}$Au + $^{197}$Au at 1.5\emph{A} GeV are nicely reproduced with the inclusion of pion potential. The strongly antiflow phenomena of $\pi^{+}$ is reduced and more consistent with the FOPI data. The more attractive $\pi^{+}$ potential leads to the in-plane emission and the elliptic flow structure is the same with the experimental data.

\textbf{Acknowledgements}
This work was supported by the National Natural Science Foundation of China (Projects No. 12175072 and No. 11722546) and the Talent Program of South China University of Technology (Projects No. 20210115).

\end{document}